\documentclass[12pt]{article}
\usepackage{amsmath,amsfonts,amsbsy}
\usepackage[dvips]{graphicx,epsfig}

\def\hybrid{\topmargin 0pt    \oddsidemargin 0pt
        \headheight 0pt \headsep 0pt
        \textwidth 6.25in       
        \textheight 9.5in       
        \marginparwidth .875in
        \parskip 0pt    \jot = 1.5ex}
\hybrid

\renewcommand{\theequation}{\thesection.\arabic{equation}} \csname
@addtoreset\endcsname{equation}{section}

\def\moth{\mathsurround=0pt}
\newdimen\zo \zo=0pt

\def\tick{\leaders\hrule height 0.5ex depth 0pt \hskip 0.5pt}
\def\upboxfill{$\moth \setbox\zo\hbox{\tick}%
  \hskip 3pt\hbox to 0pt{$\tick$\hss}\hrulefill \hbox to 7.5pt{$\tick$\hss}$}

\def\dtick{\leaders\hrule height .34pt depth 0.5ex \hskip 0.5pt}
\def\downboxfill{$\moth \setbox\zo\hbox{\dtick}%
  \hskip 2pt\hbox to 0pt{$\dtick$\hss}\hrulefill \hbox to 2pt{$\dtick$\hss}$}

\def\mso{\mathfrak{so}}
\def\msu{\mathfrak{su}}

\def\msl{\mathfrak{sl}}
\def\msp{\mathfrak{sp}}

\def\Real{{\mathbb R}}

\def\integ{{\mathbb Z}}

\def\bec{\begin{center}}
\def\ec{\end{center}}
\def\a{\alpha}  

\def\b{\beta}   
\def\c{\gamma} 

\def\d{\delta} 
\def\D{\Delta}

\def\n{\nu}

\def\s{\sigma}

\def\O{\Omega}

\def\cB{{\cal B}}

\def\cL{{\cal L}}
\def\cD{{\cal D}}

\def\cO{{\cal O}}

\def\cK{{\cal K}}

\def\cN{{\cal N}}

\def\cS{{\cal S}}
\def\cP{{\cal P}}

\def\cH{{\cal H}}

\def\cO{{\cal O}}

\def\del{\partial}

\let\la=\label

\newcommand{\eq}[1]{(\ref{#1})}

\def\be{\begin{equation}}
\def\ee{\end{equation}}
\def\bea{\begin{eqnarray}}
\def\eea{\end{eqnarray}}
\def\ba{\begin{array}}
\def\ea{\end{array}}

\def\ft#1#2{{\textstyle{{\scriptstyle #1}
\over {\scriptstyle #2}}}}

\def\scs#1{\section{\sc #1}}
\def\scss#1{\subsection{\sc #1}}
\def\scsss#1{\subsubsection{\sc #1}}

\begin{document}






\hfill{January 2009}

\vspace{20pt}

\begin{center}


{\Large\sc Anyons, Deformed Oscillator  \\ Algebras and Projectors}


\vspace{30pt}
{\sc Johan Engquist}

\vspace{5pt}

{\it\small Department of Physics, University of Oslo \\
P.O. Box 1048 Blindern, N-0316 Oslo, Norway}

\vspace{5pt}

\texttt{j.p.engquist@fys.uio.no}


\vspace{35pt}  

\end{center}

\begin{center} 

{\bf ABSTRACT}\\[3ex]

\begin{minipage}{13cm}
We initiate an algebraic approach to the many-anyon problem based on deformed oscillator algebras. The formalism utilizes a generalization of the deformed Heisenberg algebras underlying the operator solution of the Calogero problem. We define a many-body Hamiltonian and an angular momentum operator which are relevant for a linearized analysis in the statistical parameter $\nu$. There exists a unique ground state and, in spite of the presence of defect lines, the anyonic weight lattices are completely connected by the application of the oscillators of the algebra. This is achieved by supplementing the oscillator algebra with a certain projector algebra.  
\end{minipage}
\end{center}


\setcounter{page}{1}

\pagebreak


\scs{Introduction} 
In the theory of identical particles in one and two spatial dimensions there is a wider notion of statistics known as {\it fractional statistics}. In two dimensions, the associated particles are known as anyons \cite{Leinaas:1977fm,Goldin:1979ki, Wilczek:1981du}. Specifying a physical many-body problem completely usually means that in addition to defining the relevant Hamiltonian we also have to specify the statistics of the particles. In the theory of identical particles in two dimensions, however, instead of externally imposing the statistics by explicit (anti)symmetrizations, giving rise to bosonic (fermionic) wave functions, it may rather be imposed on the $N$-particle wave functions $\Psi^{(N)}$ in a geometric fashion as
\bea \la{anyonsym}
\s_{\a\b}\Psi^{(N)}&=&e^{i\pi \n}\Psi^{(N)} \ , 
\eea
where $\s_{\a\b}$ is an operator exchanging particles $\a$ and $\b$ in a counterclockwise direction.  In more mathematical terms, the $\s_{\a\b}$ operators are the generating elements of the {\it braid group} $\cB_N$. This group differs from the more familiar {\it permutation group} $\cS_N$ in that the condition that the generators must square to one is relaxed, thus making it infinite-dimensional. This geometric formulation indeed turns out to allow for a wider notion of statistics -- bosons and fermions are recovered only in particular limits, viz.~when the statistical parameter $\n$ is put to either $0$ or $1$. For non-integer $\n$, the particles are known as anyons and have fractional statistics. 

Unfortunately, solving models with the anyonic symmetry conditions \eq{anyonsym} on the wave functions is hard -- even for free anyons. The main reason for why the anyon problems are difficult is because the wave functions are {\it multivalued}, implying that the $N$-anyon Hilbert spaces are not simple tensor products of single-anyon Hilbert spaces. The problem of two anyons in a harmonic potential was understood analytically a long time ago in the seminal paper by Leinaas and Myrheim \cite{Leinaas:1977fm}. The dependence of the energy $E$ on the statistical parameter $\n$ in this particular case is linear. For three and more anyons, however, the problem becomes much more challenging since, in addition to these {\it linear states}, there also exist {\it non-linear states} with a non-linear $E(\n)$ dispersion relation. The spectrum of the low-lying non-linear states is partly known; but only from perturbative \cite{Chou:1991rg,Chou:1992rp,Karlhede:1991mw,Sporre:1991ui} and numerical \cite{Sporre:1991qm, Murthy:1991vx,  Sporre:1991pm,Mashkevich:1994me} considerations for a small number of particles. The linear wave functions, which only constitute a minor part of all wave functions for $N\ge3$, are known to be holomorphic in the particle coordinates and may be treated exactly. But in spite of this, a proper algebraic construction for these states is still lacking -- even for the simplest case of two anyons with an already intriguing weight lattice \cite{Leinaas:1977fm,Leinaas:1992ac}. By working with a standard (undeformed) oscillator construction, one is forced to introduce several ``ground states'' and to impose certain ad hoc restrictions in order to avoid producing singular wave functions which are detached from the physical spectrum (see e.g.~the review in \cite{Myrheim:1998}). In this paper, we initiate an algebraic approach to the $N$-anyon problem, based on deformed oscillator algebras, which is able to capture the linear $\n$ dependence of the exact energy and angular momentum spectra and which moreover connects the entire anyonic weight lattice so that in particular only one (proper) ground state is needed.

In Section 2 we review the existing formalism with deformed Heisenberg algebras which is relevant for the problem of identical particles in $1+1$ dimensions. In Section 3 we extend this formalism to $2+1$ dimensions and introduce a projector algebra which is needed for its construction. Throughout the article the main focus will be on two anyons ($N=2$), but the formalism also partly extends to $N$ anyons which can be considered as a linearized analysis in $\n$. In Appendix \ref{sec:1Dalg} we consider a simpler model involving an algebra with only one oscillator and find its spectrum. Appendix \ref{sec:oscapp} is devoted to an explicit construction of the oscillator states for $N=2$ and calculations of their spectra.

\scs{Deformed Algebras in One Dimension} \la{defosc1Dx}
In preparation for what follows, in this section we review the {\it exchange-operator formalism} which was put forward in Refs.~\cite{Polychronakos:1992zk, Brink:1992xr,Brink:1993sz} in order to provide an operator solution to the Calogero problem of $N$ particles. The relevant Hamiltonian is given by\footnote{Throughout the article, we use units in which $x$ and $p$ are dimensionless. We put $\hbar=1$.}
\bea \la{calham} 
H^{(N)}_{Cal}=\frac12\sum_{\a=1}^N (p_\a^2+x_\a^2)+\sum_{\a<\b}^N\frac{g}{(x_\a-x_\b)^2}\ , 
\eea 
where $x_\a$ and $p_\a$ are the coordinates and conjugate momenta of the particles and $g$ is a dimensionless coupling constant. It is known \cite{Leinaas:1988,Polychronakos:1988tm,Hansson:1991uc} that this model is related to fractional statistics of identical particles in one dimension\footnote{There also exists an alternative formulation of fractional statistics in one dimension more in the spirit of Schr\"odinger quantization \cite{Leinaas:1977fm}. This approach will not be pursued here. Note, however, that the two different notions of fractional statistics have a common origin in two dimensions as shown in \cite{Hansson:1991uc}.}. Furthermore, it serves as an effective description of the problem of $N$ anyons in a strong magnetic field, where the dynamics is restricted to the lowest Landau level  \cite{Polychronakos:1988tm,Hansson:1991uc,Ouvry:1999} (for a recent review, see Ref.~\cite{Ouvry:2007}). The coupling constant $g$ appearing in \eq{calham} is related to the statistical parameter $\n$ appearing in \eq{anyonsym} as $g=\n(\n\pm1)$ where the sign is fixed by the symmetry of the ground state (antisymmetric or symmetric). 

\scss{The Relative Motion of Two Identical Particles} \la{sec:twoid}
Let us briefly motivate the utility of the deformed oscillator algebras \cite{Vasiliev:1989qh,Vasiliev:1989re} which are part of the exchange-operator formalism. The quantum spectrum of the relative motion of two particles in one dimension is organized by the Lie algebra $\msp(2)\simeq\msl(2)$ which is defined by the commutation relations 
\bea
[J^-,J^+]=2J^0\ , \qquad \qquad [J^0,J^\pm]=\pm J^\pm\ .
\eea
The unitary representations we are interested in below belong to the {\it discrete series} of $\msl(2)$. In particular, the {\it metaplectic representations} $\cD(j)$ are specified by a lowest-weight state $|j\rangle$ with either $j=0$ (boson) or $j=\frac12$ (fermion). It is known, however, that there exist more general representations $\cD(j)$ with quantum numbers $j\in\Real$, obtained by going to the universal covering group associated with $\msl(2)$. By constructing an oscillator algebra $[a^-,a^+]=1$ out of the relative coordinate and momentum in a standard manner, the generators of the $\msl(2)$ algebra may be realized as the bilinears $J^0=\frac14\{a^-,a^+\}$ and $J^\pm=\frac12(a^\pm)^2$. The bosonic and fermionic representation spaces then consist of oscillator monomials of even and odd powers, respectively, acting on a Fock vacuum. The more general representations with $j\in\Real$, however, are {\it not} part of this oscillator construction. In order to describe these, we need the {\it deformed Heisenberg algebra} to be described in this section. As expected, the $\msl(2)$ algebra can still be represented as bilinears of these new oscillators.

Now consider the Hamitonian \eq{calham} with $N=2$. The center-of-mass motion can be separated from the relative motion and be quantized trivially. In the following we will work with the relative complex coordinates $z=z_1-z_2$, where $z_\a=(x_\a+ip_\a)/\sqrt2$. The key to solve the Calogero problem algebraically is to extract a Jastrow factor $U(z)=z^\n$ from the Calogero wave functions $\Psi(z)$. The permutation group $\cS_2$ then has a natural action on the resulting wave functions $\Phi=U^{-1}\Psi$ and an exchange-operator formalism turns out to be suitable to solve the problem. Generally speaking, an $\cS_N$ exchange operator $K_{\a\b}$, symmetric in its indices $\a, \b=1,\ldots,N$, acts on a one-particle operator $\cO_\a$ with a particle label $\a$ by conjugation as 
\bea \la{permdef}
\cO_\b=K_{\a\b}\cO_\a K_{\a\b}^{-1}\ ,
\eea
which alternatively can be written as $K_{\a\b}\cO_\a=\cO_\b K_{\a\b}$, by using that $K_{\a\b}^{-1}=K_{\a\b}$, which in turn follows from $K_{\a\b} K_{\a\b}=1$. In what follows, $K_{\a\b}$ will be referred to as a {\it permutation operator}.

Let us introduce a deformed oscillator algebra for the relative motion of the two particles which involves the statistical parameter $\n$ ($0\le\n<2$) and the permutation operator $K_{12}$:
\bea \la{defalgN2}
[a,a^\dagger]=1+2\n K \ , \qquad \qquad K\equiv K_{12}\ ,
\eea
where the relative\footnote{We will use capital letters $A_\a$ for the oscillators associated with particles $\a$, where $\a=1,\ldots,N$, and small letters for the relative oscillators $a_\a=A_\a-A_{\a+1}$.} oscillator $a=A_1-A_2$. This algebra is complete once we have specified the relations between the oscillators and the new operator $K$. Since the permutation operator acts as $KA_1=A_2K$ on a one-particle operator, as described above, we find the relations 
\bea \la{klein0}
aK+Ka=0\  \qquad\qquad {\rm and} \qquad \qquad a^\dagger K+Ka^\dagger=0\  
\eea
for the relative oscillators. An operator anticommuting with the oscillators in this way is sometimes referred to as a Klein operator or a {\it Kleinian}. 

\begin{figure}[t]
\begin{center}
\includegraphics[totalheight=.14\textheight,viewport=20 150 608 276,clip]{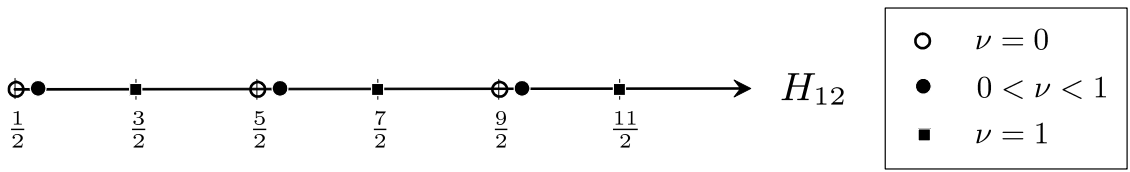}
\caption{The spectrum of the $N=2$ Calogero model for various values of $\n$. The spectrum of the permutation operator is $K|n\rangle=(-1)^n|n\rangle$. All representations may be described in the $K=1$ subspace, with an even number of oscillator excitations. In particular, $\n=0,\frac12,1$ describes the bosonic, semionic and fermionic spectrum, respectively.}
\label{fig:cal}
\end{center}
\end{figure}

We may represent the deformed oscillator algebra in terms of the relative complex coordinate $z$ defined previously as
\bea  \la{composc}
a=\nabla\equiv\del+\frac{\n}{z}(1-K)\ , \qquad\qquad a^\dagger=z\ ,
\eea 
where $\{K,z\}=0$ and $\{\del,K\}=0$. With these formulas at hand we may, after a conjugation by $U$, write the relative part of the $N=2$  Calogero Hamiltonian in an elegant way as \cite{Brink:1992xr,Brink:1993sz}
\bea
H_{12}=\frac12\{a,a^\dagger\}=a^\dagger a +\frac12+\n K\ .
\eea
It is easy to show that the oscillators $a^\dagger$ and $a$ act as raising and annihilation operators: $[H_{12},a^\dagger]=a^\dagger$ and $[H_{12},a]=-a$. By introducing a Fock vacuum $|0\rangle$, which is annihilated by $a$ and symmetric under particle exchange, so that $K|0\rangle=|0\rangle$, we immediately find the spectrum to be $H_{12}|n\rangle=(n+\frac12+\n)|n\rangle$, where we have defined $|n\rangle=(a^\dagger)^n|0\rangle$ (see Figure \ref{fig:cal}). This is the standard shifted spectrum of the Calogero model \cite{Calogero:1969ie}. In the complex coordinate representation, the wave functions take the form $\Phi_n(z)=z^n$, where $\Phi_0=1$ corresponds to the Fock vacuum. While the raising operators $a^\dagger$ in \eq{composc} act trivially on these wave functions, the lowering operator $a$ is responsible for the shift. 
In Appendix \ref{sec:1Dalg}, the algebra \eq{defalgN2} and \eq{klein0} is generalized by allowing for modified  Kleinian relations, so that the ``Calogero shift'' enters only above a particular excited state. This type of algebra will be relevant for the anyon problem which will be treated in Section \ref{sec:defalg2D}. 

\scss{$N$ Identical Particles} \la{sec:Nanyons1D}
The formalism described in the previous subsection extends straightforwardly to $N$ identical particles \cite{Polychronakos:1992zk,Brink:1992xr,Brink:1993sz}. To describe $N$ particles, however, it is more convenient to include the center-of-mass motion, which requires a total number of $2N$ oscillators $A_\a$ and $A^\dagger_\a$, with $\a=1,\ldots,N$. By using the permutation operators $K_{\a\b}$, as defined in \eq{permdef}, the $\cS_N$-{\it extended Heisenberg algebra} takes the form 
\bea \la{Nanyon1D}
&& [A_\a,A^\dagger_\b]=\d_{\a\b}\Big(1+\n \sum_{\c=1}^NK_{\a\c}\Big)-\n K_{\a\b}\ , \\
&& [A_\a,A_\b]=0\ , \qquad [A_\a^\dagger ,A_\b^\dagger]=0\ .
\eea
In a coordinate representation, these oscillators act on wave functions $\Phi(\{z_\a\})$ with a well-defined permutation symmetry, as a consequence of redefining the wave functions by the Jastrow factor $U^{(N)}(\{z_\a\})=\prod_{\a<\b}(z_\a-z_\b)^\n$. A conjugation by $U^{(N)}$ enables us to write the $N$-particle Calogero Hamiltonian \eq{calham} in the compact form
\bea \la{CalN}
\widetilde H^{(N)}_{Cal}=\frac12\sum_{\a=1}^N\{A_\a,A^\dagger_\a\}\ ,
\eea
where the oscillators can be given the complex representation
\bea  
A_\a=\del_\a+\n \sum_{\b\neq\a} \frac{1}{z_\a-z_\b}(1-K_{\a\b})\ , \qquad\qquad A_\a^\dagger=z_\a \ . 
\eea
The permutation operators act on the coordinates and derivatives in an obvious manner: $K_{\a\b}z_\a=z_\b K_{\a\b}$ and $K_{\a\b}\del_\a=\del_\b K_{\a\b}$.  By choosing the symmetric ground state wave function $\Phi_0=1$, which is annihilated by $A_\a$ and having the ground-state energy $E_0=\frac{N}{2}+\frac{N(N-1)}{2}\n$, the allowed excitations are given by symmetric combinations of the $A_\a^\dagger$ oscillators having $K_{\a\b}=1$. For details we refer to Ref.~\cite{Brink:1993sz}.

To connect to the discussion which opened Section \ref{sec:twoid}, the bilinears in the oscillators close into deformed $\n$-dependent symplectic algebras of higher rank. In addition, {\it all} the $N$-particle representations, for $N=2,3,4,\ldots$, can be combined into representations of a certain $\n$-independent higher-spin algebra which is closely related to $W_{1+\infty}$ \cite{Isakov:1995jg,Isakov:1997km}.

Incidentally, there also exist finite-dimensional (non-unitary) representations of the deformed Heisenberg algebra, as shown in Ref.~\cite{Plyushchay:1997ty} in the case of $N=2$. These representations are related to parafermions and certain generalizations thereof.


\scs{Deformed Algebras in Two Dimensions} 	    \la{sec:defalg2D}
In this section we introduce a certain $2+1$-dimensional generalization of the algebra \eq{Nanyon1D} which proves relevant for the anyon problem in two dimensions.  At this stage the algebra is conjectured, and results from a certain generalization of the one-dimensional construction. It is expected, however, to have relevance for the (linearized) analysis performed in Refs.~\cite{Chou:1992rp,Karlhede:1991mw,Sporre:1991ui}, although this remains to be spelled out in detail. Analogously to the previous section, we first formulate the two-anyon problem, for which all the features of the new formalism is present. The $N$-anyon algebra turns out to be ``linearized'' in the sense that we are able to capture the linear dependence in the $E(\n)$ dispersion relations, but not more. 

\begin{figure}[tbp]
\begin{center}
\includegraphics[totalheight=.35\textheight,viewport=4 7 525 438,clip]{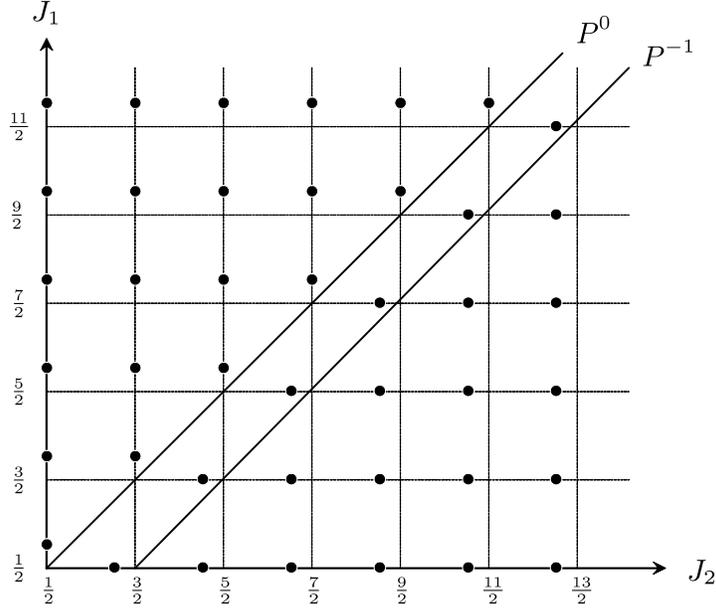}
\caption{The relative spectrum of the two-anyon problem. We have plotted the eigenvalues of $J_1$ and $J_2$, where $\cH_{12}^{(2)}=J_1+J_2$ and $\cL_{12}^{(2)}=J_1-J_2$, to emphasize the presence of defect lines around which the spectrum becomes shifted. The defect lines are associated with certain projectors $P^0$ and $P^{-1}$, as explained in Section \ref{defoscalg2}.}
\label{fig:N2}
\end{center}
\end{figure}

\scss{The Relative Motion of Two Anyons}
\scsss{The Analytic Solutions} \la{analytic}
The problem of two anyons in a harmonic potential was solved in Ref.~\cite{Leinaas:1977fm}. By using complex coordinates $z_\a=x_\a+iy_\a$ and $\bar z_\a=x_\a-iy_\a$, with $\a=1,2$, the Hamiltonian and angular momentum operator for the relative motion of the problem take the form
\bea
\cH_{12}^{(2)}=-4\del\bar\del+\frac14|z|^2\ , \qquad \quad \cL_{12}^{(2)}=z\del-\bar z\bar \del \ , 
\eea
where $z=z_1-z_2$ is the relative coordinate. In addition, we have to impose the anyonic symmetry condition \eq{anyonsym} on physical wave functions which acts as $\s_{12}: z\rightarrow e^{i\pi}z$ and $\bar z\rightarrow e^{-i\pi}\bar z$, and which is understood as a continuous interchange of particles 1 and 2 in a counterclockwise direction. We assume that $\n\ge0$; the $\n<0$ case can be treated similarly. The regular solutions to the Schr\"odinger equation $\cH_{12}^{(2)}\Psi=E\Psi$ are given by two classes of eigenfunctions
\bea
\Psi_{m,n}=\Bigg\{ 
\begin{split}
&z^\n \psi_{m,n}(z,\bar z)\ , \qquad \qquad m\ge n \\
&\bar z^{-\n} \psi_{m,n}(z,\bar z)\ , \qquad ~\quad m< n
\end{split} \qquad ,
\eea
giving rise to an interesting skew spectrum, see Figure \ref{fig:N2}. The single-valued functions $\psi_{m,n}(z,\bar z)=e^{-|z|^2/4}\sum_{p=0}^{{\rm min}(m,n)}\psi_{m,n}^pz^{m-p}\bar z^{n-p}$ are, apart from an exponential dressing, polynomials of maximal degree $(m,n)$ in the holomorphic/antiholomorphic coordinates.  

The resulting spectrum becomes
\bea \la{N2spec}
\begin{split}
& \cH_{12}^{(2)}\Psi_{m,n}=\Bigg\{ 
\begin{split}
(m+n+1+\n)\Psi_{m,n}\ , \qquad m\ge n \ , \\
(m+n+1-\n)\Psi_{m,n}\ , \qquad m< n \ ,  
\end{split} \\
& \cL_{12}^{(2)}\Psi_{m,n}=(m-n+\n)\Psi_{m,n}\ , \qquad \qquad\quad \forall ~~ m,n \ . 
\end{split}
\eea
We stress that the energy eigenvalues are shifted with respect to the $\n=0$ eigenvalues to higher or lower values, depending on whether the angular momentum is positive or negative. By restricting $m+n$ to be even, the bosonic spectrum is obtained for $\n=0$ and the fermionic spectrum for $\n=1$.

For $\n=0$ and $\n=1$, the quantum spectrum above is organized by the Lie algebra $\msp(4)$, whose ten generators are formed out of bilinears in the relative coordinates and momenta. The bosonic and fermionic representations correspond to the two {\it singleton representations} of $\msp(4)\simeq\mso(3,2)$ \cite{Dirac:1963ta}. For a detailed analysis of these representations, see e.g.~Ref.~\cite{Leinaas:1992ac}. Interestingly, the two-anyon representations with $\n\neq0,1$ are {\it not} proper $\msp(4)$ representations in general, due the presence of defect lines, but rather to a deformed version of $\msp(4)$. In Appendix \ref{sec:1Dalg} we examine a deformed $\msp(2)$ algebra which takes into account the presence of defects by containing projectors explicitly -- see \eq{sp2def}.  

\scsss{An $N=2$ Oscillator Algebra} \la{defoscalg2}
A naive oscillator construction of the two-anyon problem based on an undeformed algebra fails. As reviewed in Refs.~\cite{Leinaas:1992ac,Myrheim:1998}, due to the presence of a ``defect line'' in the anyon weight lattice, which cannot be crossed using these oscillators, there is no unique ground state. In relation to this, one is forced to impose certain ad hoc restrictions concerning the allowed oscillator combinations in order to avoid producing singular wave functions (i.e.~to avoid passing the defect line).

Therefore, we will introduce a {\it deformed oscillator algebra} for the relative motion -- an approach which was successful in one dimension as reviewed in Section \ref{defosc1Dx}. The connection to an explicit coordinate representation, however, will be postponed to a future publication. The algebra will act on {\it single-valued wave functions}, since a Kleinian is required to have a well-defined action on the eigenstates. A look at the algebra in \eq{defalgN2}, which is valid for one oscillator, suggests the following form of the ``diagonal'' part of the algebra:
\bea \la{diagalgN2}  
&& [a,a^\dagger]=1+2\n R^1_1\ , \qquad\qquad [b,b^\dagger]=1+2\n R_2^2\ , 
\eea
where the index $i=1,2$ refers to the type of oscillator. To prevent cluttering the equations, for two particles we will often suppress the particle labels on the operators -- a more proper notation for the new operators is $(R^i_i)_{12}$ as will be described in Section 3.2. The ``mixed'' parts of the algebra will necessarily involve new operators, $R_{12}\equiv (R_{12})_{12}$ and $R_1^2\equiv (R_1^2)_{12}$, together with their hermitian conjugates $R^{12}=-(R_{12})^\dagger$ and $R^{1}_2=(R_{1}^2)^\dagger$, whose application on states will change their quantum numbers. A general form of the remaining commutation relations is
\bea \la{mixalgN2}
\begin{split}
& [a,b]=2\n R_{12}\ ,~~\qquad \qquad \qquad [a^\dagger,b^\dagger]=2\n R^{12}\ , \\
& [a,b^\dagger]=2\n R_1^2\ ,  \qquad \qquad  ~~\qquad [b,a^\dagger]=2\n R_2^1 \ .
\end{split}
\eea 
Leaving these aside for the moment, our first objective is to explain the operators $R_i^i$ appearing in \eq{diagalgN2}. It turns out that only the sum of these operators amounts to the standard permutation operator which appeared in the previous section, meaning that $K=R_1^1+R_2^2$. Here, however, with the presence of two oscillators, we have the Kleinian relations $\{a,K\}=0$ and $\{b,K\}=0$ and similarly for the raising operators. To appreciate these statements we have to introduce a certain projector algebra, which will be described next. Henceforth, we will assume that $\n\ge0$; the case $\n<0$ may be treated in complete analogy.

\begin{figure}[tbp]
\begin{center}
\includegraphics[totalheight=.3\textheight,viewport=137 104 412 363,clip]{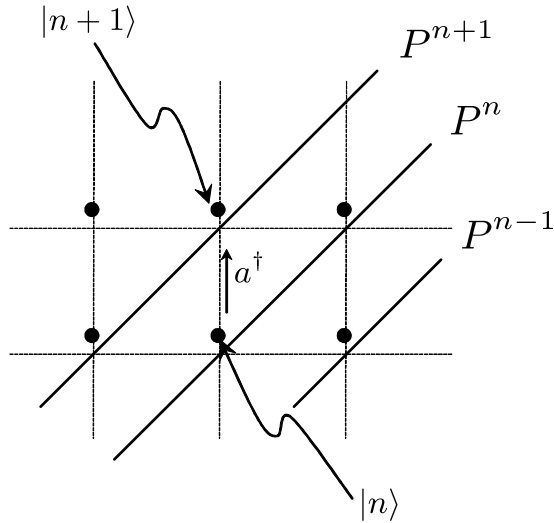}
\caption{The exchange relation $a^\dagger P^n=P^{n+1}a^\dagger$ illustrated in an example. We zoom in around states with angular momenta $n>0$ and $n+1$ in the anyon weight lattice of Figure \ref{fig:N2}. Consider the states $|n\rangle$ and $|n+1\rangle=a^\dagger|n\rangle$ with angular momenta $n$ and $n+1$, respectively. Since $a^\dagger$ raises the angular momentum by one unit, it is equivalent to $(i)$ first project with $P^n$ onto $|n\rangle$ and then act with $a^\dagger$ and $(ii)$ first act with $a^\dagger$ and then project with $P^{n+1}$. In equations this reads $a^\dagger P^n|n\rangle=a^\dagger|n\rangle=|n+1\rangle=P^{n+1}a^\dagger|n\rangle$.}
\label{fig:proj}
\end{center}
\end{figure}

It turns out that, in order to define the operators $R_i^i$ appearing in \eq{diagalgN2}, and to find their relations to the oscillators, we first need to introduce operators which project onto states with fixed relative angular momentum $\cL_{\a\b}=\cL_{12}=(\cL_{12})^\dagger$, where $\a$ and $\b$ are particle labels. For convenience, we anticipate the result that the angular momentum operator of our problem acts in a standard fashion with respect to the oscillators according to 
\bea   \la{step2L}
[\cL_{12},a^\dagger]=a^\dagger\ , \qquad \qquad  [\cL_{12},b^\dagger]=-b^\dagger\ ,
\eea 
as will be proven shortly (see \eq{step} and \eq{HLdef}) and which also can be inferred from \eq{N2spec}. This implies that whereas $a$ and $b^\dagger$ decrease the angular momentum, $b$ and $a^\dagger$ increase it. In particular, the monomial $(a^\dagger)^m(b^\dagger)^n$ carries angular momentum $\D \cL_{12}=m-n$. Let us therefore introduce a {\it projector algebra} associated with the relative motion between particles 1 and 2
\bea  \la{projalgN2} 
P^mP^n=\d_{mn}P^n\ , \qquad \qquad m,n\in\integ\ ,
\eea 
where $P^n\equiv P^n_{12}$, together with the {\it exchange relations}
\bea  \la{exchproj}
\begin{split} 
& a^\dagger P^n=P^{n+1}a^\dagger\ , \qquad\qquad b^\dagger P^n = P^{n-1}b^\dagger\ ,   \\ 
&  P^na=aP^{n+1}\ ,  ~~\qquad\qquad  P^nb = bP^{n-1}\ .
\end{split}
\eea 
We learn that $P^n$ projects onto states with a fixed angular momentum. The completeness relation reads $\sum_nP^n=1$. To see how the projectors act in a Fock space, introduce a Fock vacuum $|0\rangle$ such that $P^n|0\rangle=\d_{n,0}|0\rangle$ and with a fixed angular momentum $\cL_{12}|0\rangle=\n|0\rangle$. Then it is easy to prove that for the excited states ($\cN_{m,n}$ are calculated in Appendix \ref{sec:oscapp})
\bea \la{excstates}
|m,n\rangle=\cN_{m,n}(a^\dagger)^m(b^\dagger)^n|0\rangle \ 
\eea
we have that 
\bea  \la{projstate}
P^p|m,n\rangle=\d_{p,m-n}|m,n\rangle\ .
\eea  
Hence, $P^p$ projects onto states with a fixed relative monomial degree $m-n=p$, or equivalently, using the relations in \eq{step2L}, onto states carrying fixed relative angular momentum $\cL_{12}=p+\n$. In Figure \ref{fig:proj}, we illustrate the action of the projector algebra in an explicit example.
Incidentally, we point out that from the result \eq{projstate}, it is clear that the projectors may be realized explicitly as $P^n=\sum_{m\ge0}|m+n,m\rangle\langle m+n,m|$, which can be verified to be consistent with the definitions in \eq{projalgN2} and \eq{exchproj}. For a simpler example of a projector algebra involving a single oscillator, we refer to Appendix \ref{sec:1Dalg}.
 
With the $P^n$ projectors at our disposal, it is easy to construct operators which project onto either positive or negative angular momentum eigenstates, or equivalently onto the two different types of wave functions discussed in Section \ref{analytic}. Projectors with these properties are 
\bea  \la{Thetadefx}
(\Theta^1_1)_{\a\b}\equiv\sum_{n\ge0}P^n_{\a\b}\ , \qquad \qquad (\Theta^2_2)_{\a\b}\equiv\sum_{n<0}P^n_{\a\b}\ , 
\eea 
where we have written out the particle labels explicitly to facilitate a generalization to $N$ particles in the next subsection. The completeness relation can be expressed as $(\Theta^1_1)_{\a\b}+(\Theta^2_2)_{\a\b}=1$. From these projectors, we can now finally define the operators $R_i^i$ appearing in the algebra \eq{diagalgN2} as
\bea   \la{Rii}
(R_i^i)_{\a\b}\equiv (\Theta^i_i)_{\a\b} K_{\a\b} \ , 
\eea 
so that their sum indeed is given by the standard permutation operator $K_{\a\b}=(R_1^1)_{\a\b}+(R_2^2)_{\a\b}$, as claimed above. As already mentioned, in the $N=2$ case the particle labels will be suppressed, since there is only one independent combination, viz.~$R^i_i=(R^i_i)_{12}$. By projecting the standard Kleinian relations $\{a,K\}=0$ and $\{b,K\}=0$ with $(\Theta^i_i)_{12}$, we derive the {\it modified permutation relations} 
\bea  \la{modklein}
\begin{split}
& R^i_i a+ aR^i_i=(-1)^{i} KP^{-1} a\ , \qquad \quad \quad R^i_i b+ bR^i_i=(-1)^{i-1} KP^0 b\ , \\ 
& R^i_i a^\dagger+ a^\dagger R^i_i  =(-1)^{i}a^\dagger KP^{-1} \ , \quad \quad\quad
R^i_i b^\dagger+ b^\dagger R^i_i =(-1)^{i-1} b^\dagger KP^0 \ .
\end{split}
\eea 
These relations tell us that $R_i^i$ act as ordinary permutation operators, except in the vicinity of of the defect lines $P^0$ and $P^{-1}$, where there is a shift by $\pm1$. The $R_i^i$ spectrum is calculated in Appendix \ref{sec:oscapp}, see \eq{Riispec}. 

Next, we turn to the mixed parts of the algebra. Just as we need relations between the oscillators and the ``Kleinians'' $R_i^i$ given in \eq{modklein}, to specify the algebra completely, we also need relations between the oscillators and the new operators $R_{12}$, $R^{12}$, $R_1^2$ and $R_2^1$. It turns out that a consistent choice is
\bea  \la{modklein2}
\begin{split}
& bR_1^1-a^\dagger R_{12}+R_2^1 a=0\ , \qquad \qquad aR_2^2+b^\dagger R_{12}+R_1^2 b=0 \ , \\
& R_1^1b- R_{12}a^\dagger+aR_2^1 =0\ , \qquad \qquad R_2^2a+ R_{12}b^\dagger+bR_1^2 =0 \ ,
\end{split}
\eea 
and similar relations for their hermitian conjugates. These relations are consistent with the Jacobi identities of the algebra under consideration and in addition gives rise to natural generalized step operators; more precisely, if we define the commuting generators (``deformed'' $\msp(4)$ Cartan generators)
\bea  
&&J_1\equiv\ft12\{a,a^\dagger\}=a^\dagger a+\frac12+\n R_1^1\ , \\ 
&&J_2\equiv\ft12\{b,b^\dagger\}=b^\dagger b+\frac12+\n R_2^2\ ,
\eea 
their eigenvalues will step in a way consistent with the two-anyon weight lattice, viz. 
\bea  \la{step}
\begin{split}
& [J_1,a^\dagger ]=a^\dagger-\n a^\dagger KP^{-1}\ , \qquad\qquad [J_1,b^\dagger]=-\n b^\dagger K P^0\ , \\ 
& [J_2,b^\dagger]=b^\dagger-\n b^\dagger KP^0\ , \qquad \qquad~~~ [J_2,a^\dagger]=-\n a^\dagger KP^{-1}\ , 
\end{split}
\eea 
where we have used the relations in \eq{modklein} and \eq{modklein2}. For instance, the first of these commutation relations means that when applying $a^\dagger$ in a Fock space, the eigenvalue of $J_1$ will always step in units of $1$, except when the defect line $P^{-1}$ is crossed when the change instead is $1\pm\n$, where the sign is fixed by the action of $K$ on the ground state. This is precisely what is seen in the weight lattice, see Figure \ref{fig:N2}. The remaining commutators in \eq{step} can be given similar interpretations. 

The relations between the $R_i^i$ operators and the remaining $R$ operators are given by
\bea \la{zxb9}
R_1^1R_2^2=0\ , \qquad [R_i^i,R_{12}]=0\ , \qquad [R_i^i,R_1^2]=(-1)^iR_1^2K(P^0+P^{1})\ ,
\eea
whereas the relations between the $R_i^j$ and $R_{ij}$ operators are more involved and not very useful. For instance, we have the commutation relation $2\n[R_{12},R^{12}]=2(J_1^1R_2^2+J_2^2R_1^1)+aR_2^1b^\dagger+b^\dagger R_2^1a+bR_1^2a^\dagger+a^\dagger R_1^2b$. Let us mention that we have been unable to find an expression similar to \eq{Rii} for the $R_i^j$ and $R_{ij}$ operators, expressing them in terms of the permutation operator and the projectors. As opposed to $R_i^i$, some of these operators carry net angular momentum and can consequently not be expressed solely in terms of $K$ and $P^n$.

We are now in a position to define the Hamiltonian and angular momentum operator for the relative motion of two anyons:
\bea \la{HLdef}
\cH_{12}\equiv J_1+J_2=a^\dagger a+b^\dagger b+1+\n K\ , \qquad \qquad \cL_{12}\equiv J_1-J_2\ . 
\eea
These definitions are motivated by the fact that $(i)$ they reduce to the correct operators for $\n=0$ and $(ii)$ they give the correct one-dimensional restriction (for instance by putting $J_2=0$). Their commutation relations with the oscillators are given by \eq{step2L} and 
\bea   \la{step2}
& [\cH_{12},a^\dagger]=a^\dagger-2\n a^\dagger KP^{-1}\qquad \qquad 
& [\cH_{12},b^\dagger]=b^\dagger-2\n b^\dagger KP^0
\eea 
both of which follow from \eq{step}. The excited oscillator states defined in \eq{excstates} then obtain the energy and angular momentum eigenvalues
\bea  \la{N2spec2}
&&\cH_{12}|m,n\rangle=\Bigg\{
\begin{split}
& (m+n+1+\n)|m,n\rangle \quad\qquad m\ge n\ , \\ 
& (m+n+1-\n)|m,n\rangle \quad\qquad m< n\ ,  
\end{split} \\
&&\cL_{12}|m,n\rangle=(m-n+\n)|m,n\rangle \qquad\qquad \qquad \forall  ~  m,n \ ,
\eea 
in agreement with the analytical result in \eq{N2spec}. In this oscillator construction, the bosonic-like and fermionic-like representations with $K=1$ and $K=-1$, respectively, are combined. By restricting to the $K=1$ sector (with $m+n$ even), the bosonic spectrum is obtained for $\n=0$, the semionic spectrum for $\n=\frac12$ and the fermionic spectrum for $\n=1$. A more detailed analysis of the Fock space is presented in Appendix \ref{sec:oscapp} and an algebra which describes the various one-dimensional restrictions (for fixed $m$ or $n$) is described in Appendix \ref{sec:1Dalg}. Let us stress that in this oscillator construction, {\it all} points in the anyonic weight lattice are connected by the application of oscillators, reachable from a unique Fock ground state. This is in sharp contrast with the undeformed oscillator construction where several ``ground states'' are needed and the defect lines, described in our language by $P^0$ and $P^{-1}$, must not be crossed (see e.g.~the discussion in Ref.~\cite{Myrheim:1998}). 

\scss{$N$ Anyons}
In an attempt to generalize our algebraic construction to $N$ anyons, we immediately realize that the resulting construction will not be able to capture the full non-linear dependence in the $E(\n)$ dispersion relations, since the algebra resulting from a mere ``covariant lift'' of the results \eq{Nanyon1D}, \eq{diagalgN2} and \eq{mixalgN2} will only involve two-body interactions. Nevertheless, as a consequence of the latter and by consistency of the algebra, our model will be able to capture not only the correct linear dependence (proven in some cases where we have access to a perturbative analysis and are able to compare), but also succeeds in connecting all points in the anyon weight lattice. 

\scsss{Existing Analytic Solutions} \la{existan}
Only a tiny part of all solutions of the $N$-anyon problem is known analytically \cite{Wu:1984py,Polychronakos:1991hh, Dunne:1991bc, Grundberg:1991, Cho:1991nf, Cho:1992sb,  Mashkevich:1991ik}. The $E(\n)$ dispersion relations corresponding to these known wave functions are all linear. Let us briefly describe the simplest analytic  solutions to the problem of $N$ anyons in a harmonic potential. By redefining the wave functions by a factor of $\exp(-\frac12\sum_\a |z_\a|^2)$, the Hamiltonian and angular momentum operators read
\bea \la{HLN}
\cH^{(N)}=\sum_{\a=1}^N\big(-2\del_\a\bar\del_\a+z_\a \del_\a+\bar z_\a \bar \del_\a+1\big)\ , \qquad 
\cL^{(N)}=\sum_{\a=1}^N\big(z_\a \del_\a-\bar z_\a \bar \del_\a\big)\ ,
\eea 
where we use the complex coordinates $z_\a=x_\a+iy_\a$ and $\del_\a=\del/\del z_\a$. By taking into account the anyonic symmetry condition \eq{anyonsym}, we determine the ground state wave function to be
\bea \la{gstN}
\Psi^{(N)}_{0}=\prod_{\a<\b}^N(z_\a-z_\b)^\n\ , 
\eea 
with eigenvalues 
\bea \la{ENLN}
E_{0}^{(N)}=N+\ft{1}{2}N(N-1)\n \ , \qquad \qquad L_{0}^{(N)}=\ft{1}{2}N(N-1)\n \ .
\eea	
Excited states built upon this ground state are obtained by applying appropriate holomorphic functions ${\cal P}(\{z_\a\})$ symmetric in the particle coordinates  to the ground state. These wave functions all have linear $E(\n)$ dispersion relations $E=\c(N)+\frac{N(N-1)}{2} \n$, where $\c(N)$ is a constant. Another class of analytic solutions is built upon the state $\Psi'^{(N)}=\prod_{\a<\b}(\bar z_\a-\bar z_\b)^{2-\n}$, which has eigenvalues $E'^{(N)}=N+\frac{N(N-1)}{2}(2-\n)$ and $L'^{(N)}=\frac{N(N-1)}{2}(\n-2)$. The energy eigenvalues of the excited states built upon $\Psi'^{(N)}$ take the form $E=\c'(N)+\frac{N(N-1)}{2}(2-\n)$, where $\c'(N)$ is a constant. 
 
Concerning the {\it non-linear wave functions}, with a non-linear $E(\n)$ dispersion relation, very little is known analytically. There are, however, some numerical results for the low-lying spectrum of three and four anyons \cite{Sporre:1991qm,Murthy:1991vx,Sporre:1991pm,Mashkevich:1994me} as well as some perturbative results \cite{Karlhede:1991mw,Chou:1991rg,Chou:1992rp}. Nevertheless, certain simplifying observations may make the problem tractable; this approach will be elaborated on in a future publication \cite{wip}.   

Finally it is important to realize that the angular momentum has a $\n$ linear dependence for {\it all} states, linear as well as nonlinear -- a fact that can be proven by general arguments \cite{Mashkevich:1995ep}. 

\scsss{A Linearized Deformed Oscillator Algebra} \la{sec:defalgNx}
To describe $N$ anyons algebraically, we need $2N$ raising operators $A^\dagger_\a$ and $B^\dagger_\a$ and $2N$ annihilation operators $A_\a$ and $B_\a$. To facilitate the presentation of the algebra, we combine these into $\msu(2)$ doublets ($i=1,2$)
\bea
A_{\a i}=(A_\a,B_\a)\ , \qquad \qquad A_{\a}^i=(A^\dagger_\a,B^\dagger_\a)\ ,
\eea 
and define their associated {\it oscillator grading} 
\bea \la{oscgrad}
g(i)\equiv g(A_{i\a})\equiv g(A_{\a}^i)\ , \qquad \qquad g(i)=(-1)^{i-1} \ .
\eea
We now propose the following {\it linearized $N$-anyon algebra}
\bea \la{Nalg1}
&&  [A_{\a i},A_{\b}^j]=\d_{\a\b}\d_i^j+\n\d_{\a\b}\sum_{\c=1}^N(R_i^j)_{\a\c}-\nu (R_i^j)_{\a\b}\ ,  \\ \la{Nalg2}
&&  [A_{\a i},A_{\b j}]=\n\d_{\a\b}\sum_{\c=1}^N(R_{ij})_{\a\c}-\nu (R_{ij})_{\a\b}\ , \\ \la{Nalg3}
&& [A_\a^i,A_\b^j]=\n\d_{\a\b}\sum_{\c=1}^N(R^{ij})_{\a\c}-\nu (R^{ij})_{\a\b}\ . 
\eea 
It results from a straightforward extension of the previous results displayed in \eq{Nanyon1D}, \eq{diagalgN2} and \eq{mixalgN2} and involves the operators $(R_i^j)_{\a\b}=((R_j^i)_{\a\b})^\dagger$, $(R_{ij})_{\a\b}=-(R_{ji})_{\a\b}$ and $(R^{ij})_{\a\b}=-((R_{ij})_{\a\b})^\dagger$ which are all symmetric in the particle labels $\a$ and $\b$. We notice that the center-of-mass oscillators
\bea \la{comosc}
c_i=\frac{1}{\sqrt{N}}\sum_{\a=1}^N A_{i\a}\ , \qquad\qquad c^i=\frac{1}{\sqrt{N}}\sum_{\a=1}^N A_\a^i\ , 
\eea
obey an algebra $[c_i,c^j]=\d_i^j$ and decouples completely from the relative oscillators 
\bea \la{relosc}
 a_{i\a}=(a_\a,b_\a)=\frac{1}{\sqrt2}(A_{\a i}-A_{\a+1,i})\ , \qquad 
 a_{\a}^i=(a_\a^\dagger,b_\a^\dagger)=\frac{1}{\sqrt2}(A_{\a}^i-A_{\a+1}^i)\ ,
\eea
in the sense that $[c_i,a_{\a j}]=0$ and $[c^i,a_{\a j}]=0$. Of course, only $N-1$ of the $a_{\a i}$ oscillators are independent, since $\sum_{\a=1}^N a_{\a i}=0$. The two-anyon algebra \eq{diagalgN2} and \eq{mixalgN2} is recovered if we express the algebra in \eq{Nalg1}--\eq{Nalg3} for $N=2$ in terms of the relative oscillators \eq{relosc}. The $N$-particle algebra \eq{Nanyon1D} in one dimension is recovered if we focus on the $i=1$ content of the same.

The permutation operators $K_{\a\b}$ are defined by 
\bea \la{stdpermN}
K_{\a\b}A_{\a i}=A_{\b i}K_{\a\b}\ , \qquad \quad K_{\a\b}A_{\a}^i=A_{\b}^i K_{\a\b}\ , \qquad \quad  K_{\a\b}K_{\a\b}=1\ ,
\eea
and are related to the operators $R_i^i$ appearing in \eq{Nalg1} as 
\bea \la{modkleinN}
K_{\a\b}=\sum_{i=1}^2 (R_i^i)_{\a\b} \ ,
\eea
in exact analogy with the $N=2$ case, recall \eq{Thetadefx} and \eq{Rii}. In the $N$-particle case, however, we need $N(N-1)/2$ projectors $P^n_{\a\b}$ which are interpreted similarly as in Section \ref{defoscalg2}: they project onto states having relative angular momentum $\cL_{\a\b}=n+\n$ between particles $\a$ and $\b$. Using the oscillator grading \eq{oscgrad}, the relevant two-body projector algebra can be written covariantly as
\bea
&& P^m_{\a\b}P^n_{\a\b}=\d_{mn}P^n_{\a\b}\ , \\
&& P^n_{\a\b}(A_{i\a}-A_{i\b})=(A_{i\a}-A_{i\b})P^{n+g(i)}_{\a\b}\ , \\
&& P^n_{\a\b}(A_{\a}^i-A_{\b}^i)=(A_{\a}^i-A_{\b}^i)P^{n-g(i)}_{\a\b}\ ,
\eea
whereas the two-body center of mass decouples $[P^n_{\a\b},A_{i\a}+A_{i\b}]=0$. We also define $P_{\a\b}^nA_\c=A_\c P_{\a\b}^n$ for $\a\neq\b\neq\c$. The modified permutation relations are obtained by acting on \eq{stdpermN} with the projector $(\Theta^i_i)_{\a\b}$ defined in \eq{Thetadefx}:
\bea \la{modepermzz}
&& \{ (R_i^i)_{\a\b},A_{j\a}-A_{j\b}\}=(-1)^{i}g(j)K_{\a\b}P_{\a\b}^{j-2}(A_{j\a}-A_{j\b})\ , \\ 
&& \{ (R_i^i)_{\a\b},A_{\a}^j-A_{\b}^j\}=(-1)^{i}g(j)(A_{\a}^j-A_{\b}^j)K_{\a\b}P_{\a\b}^{j-2}\ . 
\eea
By summing over $i$, the relations in \eq{modepermzz} clearly implies that $ \{K_{\a\b},A_{j\a}-A_{j\b}\}=0$. The relations between $(R_i^j)_{\a\b}$, with $j\neq i$, and the oscillators read
\bea
&& A_{i\a}(R_j^i)_{\a\b}+(R_i^i)_{\a\b}A_{j\a}+(R_{ij})_{\a\b}A^i_{\b}=0 \ ,\\
&& (R_j^i)_{\a\b}A_{i\a}+A_{j\a}(R_i^i)_{\a\b}+A^i_{\b}(R_{ij})_{\a\b}=0 \ ,
\eea
and similar relations for their hermitian conjugates. These relations are consistent with the Jacobi identities. The two-body center-of-mass oscillators $A_{j\a}+A_{j\b}$ decouple from the $R$ operators, so that for example $[(R_{ij})_{\a\b} ,A_{j\a}+A_{j\b}]=0$. Finally, the relations between the $R$ operators are similar to those in \eq{zxb9}.

We define the $N$-particle Hamiltonian and angular momentum operators as
\bea \la{HN}
 && \cH\equiv \frac12\sum_{i,\a}\{A_{i\a},A_\a^i\}=\sum_{i,\a}\Big(A_\a^iA_{i\a}+\frac12\Big)+\n\sum_{\a<\b}K_{\a\b}\ , \\ \la{LN}
 && \cL\equiv \frac12\sum_{i,\a}g(i)\{A_{i\a},A_\a^i\}=\sum_{i}g(i)\Big(\sum_{\a} 
 A_\a^i  A_{\a i}+\n\sum_{\a<\b}(R_i^i)_{\a\b} \Big) \ , 
\eea
where the sums run over $i=1,2$ and $\a=1,\ldots,N$. Again, these definitions are motivated by the fact that they reduce to the correct expressions for $\n=0$ and furthermore give the correct one-dimensional restrictions, cf.~\eq{CalN} in Section \ref{sec:Nanyons1D}. A straightforward, calculation gives the results 
\bea   \la{stepHN}
&& [\cH,A_{\a}^i]= A_{\a}^i-\n\sum_{\b\neq\a}(A^i_{\a}-A^i_{\b})K_{\a\b}P^{i-2}_{\a\b} \ , \\ \la{stepLN}
&& [\cL,A^i_{\a}]=g(i)A^i_{\a}\ ,  \la{stepLNx}
\eea
from which we see that $A_\a^i$ act as ordinary raising operators, except that they will pick up extra contributions when passing some of the defect lines $P_{\a\b}^{i-2}$.  It follows from \eq{stepHN} that the center-of-mass oscillators defined in \eq{comosc} obey $[\cH,c^i]=c^i$, as we expect. 

To find the spectrum of the model, we introduce a lowest-weight state $|0\rangle$ with fixed $K_{\a\b}$ and $P^n_{\a\b}$ eigenvalues. We choose a symmetric ground state which is invariant under arbitrary exchanges of pairs of particles: 
\bea \la{lwsN}
A_{i\a}|0\rangle=0\ , \qquad \qquad K_{\a\b}|0\rangle=|0\rangle\ , \qquad \qquad  P^n_{\a\b}|0\rangle=\d_{n,0}|0\rangle\ . 
\eea
The $K_{\a\b}=1$ and $K_{\a\b}=-1$ sectors built upon this ground state correspond to starting from the bosonic and fermionic ends of the spectra at $\n=0$, respectively. For the $N$-particle ground state $|0\rangle$, we directly find from \eq{HN} and \eq{LN} the correct eigenvalues \eq{ENLN} that was calculated previously; to show this we use that $(R_i^i)_{\a\b}|0\rangle=\d_{i}^1|0\rangle$, which results after applying the projector $(\Theta_i^i)_{\a\b}$ defined in \eq{Thetadefx} to the second equation in \eq{lwsN}. In a coordinate representation, the oscillators under consideration act properly on single-valued wave functions, so that a Jastrow factor of the form $U=\prod_{\a<\b}(z_\a-z_\b)^\n$ has to be extracted from the multi-valued wave functions analogously to the one-dimensional case, cf.~Section \ref{sec:Nanyons1D}. In particular, this means that the ground state $|0\rangle$ should correspond to the single-valued wave function ${\it \Phi}_{0}^{(N)}=1$. 

Let us construct the lowest-lying three-particle oscillator states. For these states we have access to both numerical and perturbative results, to which we can compare.  We want to find states $|\Psi\rangle$ which simultaneously diagonalize the Hamiltonian and the angular momentum operators appearing in \eq{HN} and \eq{LN}, viz.
\bea
\cH|\Psi\rangle=E|\Psi\rangle \ ,\qquad \qquad \cL|\Psi\rangle=L|\Psi\rangle\ ,
\eea
and which in addition are either totally symmetric or antisymmetric under any interchange of particle labels (so that $K_{\a\b}=\pm1$). As shown before, the center-of-mass excitations $|(n_i)\rangle=\prod_{i=1}^2(c^i)^{n_i}|0\rangle$ decouple from the relative excitations and therefore all have the same $\n$ dependence as the ground state. In the following, we will focus on the relative excitations with oscillators $a^i_\a=(a_\a^\dagger,b_\a^\dagger)$ defined in \eq{relosc}. Recall from the discussion in Section \ref{existan} that there are two classes of states for $N\ge3$: the linear and the non-linear ones. For $N=3$, all linear states have energies going as $3\n$. We will refer to the states constructed below as either linear or ``non-linear'', although the latter ones will capture only the linear dependence on $\n$ in a sense described below; this is due to the limitations of the current formulation of our model. On the other hand, since the angular momentum operator has a standard commutation relation with the oscillators, see \eq{stepLNx}, {\it all} of its eigenvalues come out correctly compared to an exact analysis; the $\n$ dependence is fixed by the $\n$ dependence of ground state.  

As already mentioned, for the ground state $|0\rangle$, we immediately read off from \eq{HN} and \eq{LN} that $E_{0}=3+3\n$ and $L_{0}=3\n$. In the following we label states by their energy and angular  momentum eigenvalues for $\n=0$ so that for the ground state we have that $|3,0\rangle=|0\rangle$. There are no single-oscillator excitations but there are three states involving two relative oscillators which explicitly are given by (unnormalized)
\bea
&& |5,2\rangle=(a^\dagger_1a^\dagger_1+a^\dagger_2a^\dagger_2+a^\dagger_3a^\dagger_3)|0\rangle\ , \\ 
&& |5,0\rangle=(a_{1}^\dagger b_{1}^\dagger+a_{2}^\dagger b_{2}^\dagger +a_{3}^\dagger b_{3}^\dagger )|0\rangle\ , \\
&& |5,-2\rangle=(b_{1}^\dagger b_{1}^\dagger+b_{2}^\dagger b_{2}^\dagger +b_{3}^\dagger b_{3}^\dagger )|0\rangle\ , 
\eea
with eigenvalues summarized in Table \ref{table:E5}. As described above, the angular momentum eigenvalues come out correctly compared to an exact analysis. The energy eigenvalues of the linear states $|5,2\rangle$ and $|5,0\rangle$ come out correctly  and the linear $\n$ dependence is absent for the ``non-linear'' state $|5,-2\rangle$. The latter result is in agreement with the perturbative analysis of Refs.~\cite{Sporre:1991ui,Chou:1992rp}, where it was shown that the energy of the non-linear state goes as $E(\n)=5+\frac92\n^2\ln\frac43+\cO(\n^3)$. 
\begin{table}[tdp]
\begin{center}
\begin{tabular}{|c|c|c|}
\hline 
State  & $E$ & $L$ \\
\hline \hline
$|5,2\rangle$ & $5+3\n$ & $2+3\n$ \\
\hline 
$|5,0\rangle$ & $5+3\n$ & $3\n$ \\
\hline 
$|5,-2\rangle$  & $5$ & $-2+3\n$ \\ 
\hline 
\end{tabular}
\end{center}
\caption{Energy and angular momentum eigenvalues of states with two relative oscillator excitations. The states $|5,2\rangle$ and $|5,0\rangle$ correspond to linear states while $|5,-2\rangle$ corresponds to a ``non-linear" state.}
\label{table:E5}
\end{table}

For three relative oscillator excitations we find the diagonal states
\bea
&&|6,3\rangle=(a_{2}^\dagger-a_{3}^\dagger)(a_{3}^\dagger-a_{1}^\dagger)(a_{1}^\dagger-a_{2}^\dagger)|0\rangle\ , \\
&& |6,1\rangle=(b_{2}^\dagger-b_{3}^\dagger)(a_{3}^\dagger-a_{1}^\dagger)(a_{1}^\dagger-a_{2}^\dagger)|0\rangle+c.p.\ , \\
&& |6,-1\rangle=(a_{2}^\dagger-a_{3}^\dagger)(b_{3}^\dagger-b_{1}^\dagger)(b_{1}^\dagger-b_{2}^\dagger)|0\rangle+c.p.\ , \\
&&|6,-3\rangle=(b_{2}^\dagger-b_{3}^\dagger)(b_{3}^\dagger-b_{1}^\dagger)(b_{1}^\dagger-b_{2}^\dagger)|0\rangle\ , 
\eea
where $c.p.$~stands for cyclic permutations. The associated eigenvalues are summarized in Table \ref{table:E6}. The states $|6,3\rangle$ and $|6,1\rangle$ are linear while $|6,-1\rangle$ and $|6,-3\rangle$ are ``non-linear''. While the angular momentum eigenvalues all come out correctly, only the linear $\n$ dependence of the exact energy eigenvalues are captured. The perturbative analyses in Refs.~\cite{Sporre:1991ui,Chou:1992rp} shows that the energy eigenvalue of the non-linear state $|6,-3\rangle$ is $E(\n)=6-\frac{3}{2}\n+\frac{9}{8}\n^2(3\ln\frac43-1)+\cO(\n^3)$, in agreement with the result in Table \ref{table:E6} to linear order. As also seen in the table, the linearized energy of the state $|6,-1\rangle$ is $E=6+\frac32 \n$, in harmony with the result of Ref.~\cite{Sporre:1991ui}. 
\begin{table}[tdp]
\begin{center}
\begin{tabular}{|c|c|c|}
\hline 
State & $E$ & $L$ \\
\hline \hline
$|6,3\rangle$ & $6+3\n$ & $3+3\n$ \\
\hline 
$|6,1\rangle$ & $6+3\n$ & $1+3\n$ \\
\hline 
$|6,-1\rangle$  & $6+\frac{3\n}{2}$ & $-1+3\n$ \\ 
\hline
$|6,-3\rangle$  & $6-\frac{3\n}{2}$ & $-3+3\n$ \\ 
\hline 
\end{tabular}
\end{center}
\caption{Energy and angular momentum eigenvalues of states with three relative oscillator excitations. The states $|6,3\rangle$ and $|6,1\rangle$ correspond to linear states while $|6,-1\rangle$ and $|6,-3\rangle$ correspond to ``non-linear" states.}
\label{table:E6}
\end{table}

We emphasize that one may equally well start from the fermionic end with states having $K_{\a\b}=-1$. The first fermionic ``non-linear" state is given by 
\bea
|5,0\rangle_F=(A^\dagger_1-A^\dagger_2)B^\dagger_3|0\rangle+c.p.
\eea
One can check that for this state, the energy correction vanishes to linear order in $\n$, in agreement with the result in Refs.~\cite{Chou:1991rg,Sporre:1991ui}.

Consequently, for the lowest-lying three-anyon states, we find agreement with the exact anyon spectrum to linear order in $\n$. This can be summarized in the equations
\bea
E(\n)=E_{linear}(\n)+\cO(\n^2)\ , \qquad \qquad L(\n)=L_{linear}(\n)\ ,
\eea
where $E(\n)$ and $L(\n)$ are the results obtained from numerics and $E_{linear}$ and $L_{linear}$ are the eigenvalues obtained from the model defined by \eq{HN} and \eq{LN} as well as from perturbative analyses. We expect this agreement to hold for arbitrary $N$, although this needs to be elaborated on further. It is a quite tedious, but straightforward, exercise to determine the eigenvalues of an arbitrary state, due to the presence of the projectors and the modified permutation operators in the algebra.

\scs{Conclusions and Outlook}
In this paper we have examined an $N$-particle model defined by the Hamiltonian and angular momentum operator in \eq{HN} and \eq{LN} together with the deformed oscillator algebra \eq{Nalg1} -- \eq{Nalg3}. The algebra is nonstandard due to the presence of defect lines $P_{\a\b}^n$ which implies that the standard permutation relations need to be modified. We find step operators which act in a nearly standard fashion -- the difference being that extra contributions proportional to $\n$ are picked up whenever some defect line is passed. 

Since the exact anyon angular momentum spectrum is linear in $\n$, our model completely reproduces it. Moreover, the energy \eq{ENLN} of the $N$-anyon ground state comes out correctly. For the lowest-lying three-anyon states, we have shown that the model is able to capture the linear $\n$ dependence of the exact energy eigenvalues $E(\n)$, for both the linear and non-linear states; the spectrum of the linear states is thus fully reproduced. Indeed, the algebra considered in this paper is of direct relevance for a perturbative analysis in $\n$ along the lines of Refs.~\cite{Chou:1992rp,Karlhede:1991mw,Sporre:1991ui}, where it was shown how to calculate the linear and quadratic parts of the $N$-anyon spectrum. In lack of a complete classification of the first-order perturbative spectrum in $\n$, we are unable to compare our resulting spectrum further. Nevertheless, we expect that the agreement will continue to hold. 

To develop the algebra further, by including contributions of non-linear nature, it is crucial to make contact with the coordinate representation analogously to the one-dimensional case, as described in Section \ref{defosc1Dx}. There, the connection can be made explicit by utilizing a coherent-state representation \cite{leinaas1996}. We expect that a similar construction should work in the two-dimensional case, although it is expected to be more involved. Ultimately, there is need for a better analytical understanding of the non-linear wave functions. In a future contribution \cite{wip}, we will describe a systematic approach to find them. At present, it is not clear whether a non-linear completion of the linearized $N$-anyon algebra \eq{Nalg1} will involve additional modifications on the right-hand side or whether it is required to modify the projector algebra, for instance by taking into account three-body interactions. We believe that the approach considered in this paper, working algebraically and analytically side by side, will be useful in gaining further insight into the many-anyon problem.

\subsection*{Acknowledgments}
The author would like to thank T.H.~Hansson, J.~Suorsa and S.~Viefers for discussions and especially J.M.~Leinaas for many interesting discussions and comments on the manuscript. 

\begin{appendix}
\renewcommand{\theequation}{\Alph{section}.\arabic{equation}}

\section{A Generalization of the $\cS_2$-extended Heisenberg Algebra}   	\la{sec:1Dalg}
In this appendix, we generalize the deformed Heisenberg algebra described in Section \ref{sec:twoid}. These algebras appear as subalgebras of the linearized $N$-anyon algebra of Section \ref{sec:defalg2D}. We keep the same structure of the commutation relations \eq{defalgN2}, 
\bea \la{algappx}
[a,a^\dagger]=1+2\n \cK^M\ ,
\eea
but modify of the permutation relations \eq{klein0}, which is signified on the operator $\cK^M$ by a superscript $M$. More specifically, we introduce a projector algebra with the properties 
\bea \la{projx}
\cP^{m} \cP^{n}=\d_{mn}\cP^{n}\ , \qquad (\cP^{n})^\dagger=\cP^{n}\ , \qquad a\cP^{n}=\cP^{n-1} a\ , \qquad a^\dagger \cP^{n}=\cP^{n+1} a^\dagger\ ,  
\eea
where $n$ is an integer, and modify the permutation relations \eq{klein0} according to  
\bea \la{klein1}
a\cK^M+\cK^M a=-\cP^{M}Ka\  \qquad\quad {\rm and} \qquad \quad a^\dagger \cK^M+\cK^M a^\dagger=-a^\dagger K\cP^{M}\ , 
\eea
for a {\it fixed} integer $M$. Here, $K=K_{12}$ is the standard permutation operator \eq{klein0}. A Fock space vacuum $|0\rangle$ is fixed by specifying the action of $\cK^M$ and $\cP^n$ ($M>0$):
\bea
a|0\rangle=0\ , \qquad \qquad \cK^M|0\rangle=0\ , \qquad \qquad \cP^{n}|0\rangle=\d_{n,0}|0\rangle\ .
\eea 
Define $\s$ to be $+$ or $-$ depending on whether $M$ is even or odd. For the (normalized) states
\bea
|n\rangle=\cN^\s_n(a^\dagger)^n|0\rangle\ , \qquad \qquad \cN^\s_n=\prod_{m=1}^n\frac{1}{\sqrt{m+2\n(-1)^{M+1}(p_M)_m^{\s}}} \ ,
\eea
we then find that the ``Calogero shift" by $\n$ will appear for excitations above the state $|M\rangle$, but not for excitations below it (the coefficients $(p_M)_m^\pm$ are defined below in \eq{projxww}). To see this, note that the $\cP^M$ and $\cK^M$ spectra are given by $\cP^M|n\rangle=\d_{n,M}|n\rangle$ and
\bea
\cK^M|n\rangle=\Bigg\{
\begin{split}
& 0 \ , ~\qquad \qquad\quad n\le M\\
& (-1)^n|n\rangle\ , \qquad n> M 
\end{split} \quad  ,
\eea
so that that, effectively, the algebra \eq{algappx} appears to be undeformed for $n\le M$. The spectrum of the ``Hamiltonian''  $H_{(M)}=\frac12\{a,a^\dagger\}=a^\dagger a +\frac12+\n\cK^M$ is found to be 
\bea
H_{(M)}|n\rangle=\Bigg\{
\begin{split}
& (n+\ft12)|n\rangle \ ,\qquad \qquad\qquad\qquad n\le M\\
& \big(n+\ft12+(-1)^{M+1}\n\big)|n\rangle\ , \qquad \, n> M 
\end{split} \quad ,
\eea
using that $[H_{(M)},a^\dagger]=a^\dagger-\n a^\dagger K\cP^M$. Incidentally, to get a shift by $(-1)^M\n$, we change sign on the right-hand-side of \eq{klein1}. Let us consider the case when  $M$ is odd; the case when $M$ is even is completely analogous. The oscillators act on normalized states as 
\bea 
&&a|n\rangle=\sqrt{n+2\n (p_M)^-_{n}}|n-1\rangle \ , \\ 
&&a^\dagger|n\rangle=\sqrt{n+1+2\n (p_M)^+_{n}}|n+1\rangle\ , 
\eea
where $(p_M)^\pm_{n}$ are the eigenvalues of the projectors $\pi^\pm_M\equiv \ft12(1\pm K)\Theta^M$ so that $\pi^\pm_M|n\rangle=(p_M)^\pm_{n}|n\rangle$. They take the values
\bea \la{projxww}
&&(p_M)^\pm_{n}=\ft12\big(1\pm(-1)^{n}\big)(\Theta^M)_{n}
=\Bigg\{ 
\begin{split}
& 0\ , \qquad \qquad \qquad \quad \,\, n\le M \  \\
& \ft12\big(1\pm(-1)^{n}\big)\ , \qquad n>M \  
\end{split} \ \quad , 
\eea
where $(\Theta^M)_{n}$ are the eigenvalues of the projector $\Theta^M=\sum_{n\ge M+1}\cP^n$ on the states $|n\rangle$.  $\Theta^M$ is an operator projecting onto states above the ``defect'' $\cP^M$. Let us point out that the modified permutation relations \eq{klein1} can be derived starting from the definition
\bea
\cK^M=\Theta^M K\ ,
\eea
and projecting the standard permutation relations in \eq{klein0} by $\Theta^M$. This is in complete analogy with the  method described in Section \ref{defoscalg2}.

With the unimodular Fock states $|n\rangle$ available, together with their duals $\langle n|=(|n\rangle)^\dagger$,  the oscillators and the projectors can be realized as (for $M$ odd)
\bea
&& a=\sum_{n=0}^\infty\sqrt{n+2\n (p_M)^-_n}|n-1\rangle\langle n| \ ,   \\
&& a^\dagger=\sum_{n=0}^\infty\sqrt{n+1+2\n (p_M)^+_n}|n+1\rangle\langle n| \ , \qquad \quad
\cP^{n}=|n\rangle\langle n|\ ,
\eea
fulfilling \eq{algappx}--\eq{klein1}. 

Interestingly, the representations of the algebra \eq{algappx}--\eq{klein1} are not proper $\msp(2)\simeq\msl(2)\simeq\mso(2,1)$ representations due to the presence of a defect. Rather, they are representations of a deformed algebra $\msl(2)_{(\n;M)}$ which is specified by two parameters. The algebra takes the form
\bea  \la{sp2def}
\begin{split}
& [J^-,J^+]=2J^0+\n J^0K(\cP^{M+1}-\cP^M)-\ft{1}{4}\n K(\cP^M+\cP^{M+1})-\ft{1}{2}\n^2\cP^{M+1}\ , \\ 
&  [J^0,J^+]=J^++\ft{1}{2}\n J^+K(\cP^{M-1}-\cP^M) \ , \\
&  [J^0,J^-]=-J^-+\ft{1}{2}\n(\cP^M-\cP^{M-1})KJ^- \ , \\
& [J^0,\cP^M]=0\ , \qquad~~~ J^\pm \cP^M=\cP^{M\pm2}J^\pm\ , 
\end{split}
\eea
together with $[\cP^n,K]=0$, $[J^0,K]=0$ and $[J^\pm,K]=0$. As expected, the generators may be realized in terms of the oscillators appearing in \eq{algappx} as
\bea
J^-=\frac12a^2\ , \qquad \qquad J^+=\frac12(a^\dagger)^2\ , \qquad \qquad J^0=\frac12H_{(M)}\ ,
\eea
which guarantees that the Jacobi identities are obeyed. One can check that there are two types of lowest-weight states $|\O\rangle$ which fulfill $J^-|\O\rangle=0$. These are realized in the Fock basis by $|\O_1\rangle=|0\rangle$ and $|\O_2\rangle=a^\dagger|0\rangle$. The representation spaces built upon these with $J^+$ consist of an even and odd number of oscillators on the Fock vacuum, respectively. There is no standard quadratic Casimir operator commuting with all the generators. All we can require is that the operator ${\cal C}_2=\frac12(J^+J^-+J^-J^+)-J^0J^0$ takes constant values on both sides of the defect. 

\section{Representation Theory for $N=2$}   	\la{sec:oscapp}
In this appendix, we present the representation theory of the two-anyon algebra explicitly using our oscillators. We are interested in representing the algebra \eq{diagalgN2} and \eq{mixalgN2}  unitarily in a Fock space. This means that the representations necessarily are infinite-dimensional. We introduce a unimodular lowest-weight state $|0\rangle$ which is annihilated by the lowering operators $a$ and $b$. To characterize the representation uniquely, we have to specify the action of the Kleinian $K$ and the projection operators $P^n$ on the ground state. We choose the conditions
\bea
K|0\rangle=|0\rangle\ , \qquad \qquad P^n|0\rangle=\d_{n,0}|0\rangle\ ,
\eea  
which state that the ground state is invariant under an exchange of the particles and that it is carrying angular momentum $\cL_{12}=\n$ (recall that $P^n$ projects onto states carrying $\cL_{12}=n+\n$). There is freedom, however, to choose alternative representations, for instance one with an odd ground state, such that $K|0\rangle=-|0\rangle$.

Next, we consider the excited states. We stress that the ordering of the oscillators is important since they do not commute. For instance, as a consequence of the commutator $[a^\dagger,b^\dagger]=2\n R^{12}$, it is clear that $b^\dagger a^\dagger |0\rangle\neq a^\dagger b^\dagger |0\rangle$. We choose a ``normal ordering'' in which the $a^\dagger$ oscillators are put to the left of the $b^\dagger$ oscillators, so that we may define  
\bea
|m,n\rangle\equiv \cN_{m,n}(a^\dagger)^m(b^\dagger)^n|0\rangle\ , \qquad \qquad \langle m,n|\equiv \cN_{m,n}\langle 0|b^na^m\ ,
\eea
where $\cN_{m,n}$ are normalization constants. By defining the norm of the ground state $\langle0|0\rangle=1$ and choosing the normalization constants to be real, we establish that
\bea
 \cN_{m,n}=\prod_{k=1}^m\frac{1}{\sqrt{k+2\n(p_1^1)^-_{k,n}}}\prod_{l=1}^n\frac{1}{\sqrt{l-2\n(p_2^2)^+_{0,l}}}\ .
\eea
Here, $(p^i_i)^\pm_{m,n}$ are the eigenvalues of the projectors $(\pi^i_i)^\pm\equiv \ft12(1\pm K)\Theta_i^i$, with $i=1,2$, so that $(\pi_i^i)^\pm|m,n\rangle=(p_i^i)^\pm_{m,n}|m,n\rangle$. They take the values
\bea
&&(p_i^i)^\pm_{m,n}=\ft12\big(1\pm(-1)^{m+n}\big)(\Theta^i_i)_{m,n}
=\Bigg\{ 
\begin{split}
& \ft12\d_i^1\big(1\pm(-1)^{m+n}\big)\ , \qquad m\ge n   \\
& \ft12\d_i^2\big(1\pm(-1)^{m+n}\big)\ , \qquad m<n  
\end{split}  \quad ,
\eea
where $(\Theta_i^i)_{m,n}$ are the eigenvalues  $\Theta_i^i|m,n\rangle=(\Theta_i^i)_{m,n}|m,n\rangle$ of an operator which projects onto states with positive ($m\ge n$) or negative  ($m<n$) angular momentum; see its  definition in \eq{Thetadefx}. Here, $\d_i^j$ is a Kronecker delta function which we allow ourselves to use in a non-covariant fashion. 

Let us mention that starting from the Kleinian $K$, we may define the projector 
\bea
\Pi^\pm\equiv\ft12(1\pm K)\ , \qquad \qquad \Pi^\pm\Pi^\pm=1\ , \qquad \qquad \Pi^\pm\Pi^\mp=0\ , 
\eea
with corresponding eigenvalues $P^\pm_{m,n}=\ft12(1\pm(-1)^{m+n})$. Its relation to the projectors above is simply $\Pi^\pm=(\pi_1^1)_\pm+(\pi_2^2)_\pm$.

The  oscillators act on the normalized states as 
\bea 
&&a|m,n\rangle=\sqrt{m+2\n (p^1_1)^-_{m,n}}|m-1,n\rangle \ , \\ 
&&a^\dagger|m,n\rangle=\sqrt{m+1+2\n (p^1_1)^+_{m,n}}|m+1,n\rangle\ , \\
&&b|m,n\rangle=\sqrt{n-2\n (p_2^2)^+_{m,n}}|m,n-1\rangle\ , \\
&&b^\dagger|m,n\rangle=\sqrt{n+1-2\n (p_2^2)^-_{m,n}}|m,n+1\rangle\ . 
\eea

Finally, for completeness, we examine the $R_i^i$ eigenvalues. From the modified permutation relations \eq{modklein} it is easy to show that 
\bea \la{Riispec}
R_i^i|m,n\rangle=\Bigg\{
\begin{split}
& \d_i^1(-1)^{m+n}|m,n\rangle\ ,\qquad  m\ge n  \\
& \d_i^2(-1)^{m+n}|m,n\rangle\ ,\qquad m<n
\end{split}  \quad .
\eea
Hence, above the defect line $P^{-1}$, i.e.~for $m\ge n$, we have that $R_2^2=0$ which means that in this upper wedge $b$ and $b^\dagger$ act as an ordinary (undeformed) oscillators, cf.~\eq{diagalgN2}.  This is exactly what is seen in the weight lattice in Figure \ref{fig:N2}: above the defect line, the $J_2$ eigenvalues are {\it not} shifted. The opposite is true below the defect, where the $a$ and $a^\dagger$ oscillators act as though they were undeformed, and the $J_1$ eigenvalues do not become shifted.
 
\end{appendix}

\end{document}